\documentclass[twocolumn,prl,showpacs,preprintnumbers,amsmath,amssymb]{revtex4}

\usepackage{graphicx}
\usepackage{dcolumn}
\usepackage{bm}

\begin{document}


\title{Multiband-Driven Superfluid-Insulator Transition of Fermionic Atoms in Optical Lattices: A Dynamical Mean-Field-Theory Study}

\author{Takuji Higashiyama}
\author{Kensuke Inaba}%
\affiliation{Department of Applied Physics, Osaka University, Suita, Osaka 565-0871, Japan}

\author{Sei-ichiro Suga}
\affiliation{Department of Applied Physics, Osaka University, Suita, Osaka 565-0871, Japan}

\date{\today}

\begin{abstract}
The superfluid-insulator transitions of the fermionic atoms in optical lattices are investigated by the two-site dynamical mean-field theory. It is shown that the Mott transition occurs as a result of the multiband effects. The quasiparticle weight in the superfluid state decreases significantly, as the system approaches the Mott transition point. By changing the interaction and the orbital splitting, we obtain the phase diagram at half filling. The numerical results are discussed in comparison with the effective boson model. 
\end{abstract}

\pacs{03.75.Lm, 05.30.Fk, 73.43.Nq}
\maketitle



Since the superfluid of trapped atomic Fermi gases $\rm^{40}K$ and $\rm^{6}Li$ was observed \cite{Regal2004,Bartenstein2004,Zwierlein2004,Kinast2004,Bourdel2004,Partridge2005}, intense theoretical and experimental studies have been done for ultracold atomic Fermi gases. In these experiments, magnetic-field Feshbach resonances provide the means for controlling both the strength of the interaction between fermionic atoms and its sign \cite{Chen2005}. These tunable interactions enable us to observe the crossover between the BCS superfluid for the weak attractive interaction and the Bose-Einstein condensate (BEC) of bound pairs for the strong attractive interaction. Furthermore, by loading atoms into optical lattices, diverse interaction configurations can be introduced. The combination of these two experimental techniques plays an important role in the study of ultracold fermionic atoms and offers the experimental description of various intriguing quantum many-body phenomena.


Recent experiments revealed fascinating aspects of fermionic atoms in three dimensional optical lattices \cite{Kohl2005,Chin2006}. 
In the ETH experiment, a band insulator in the lowest band was produced, i.e. two atoms in different hyperfine states occupy the lowest state per lattice site \cite{Kohl2005}. Controlling the interaction, they further observed the partially populated higher bands.
In the MIT experiment, a superfluidity of fermionic atom pairs in an optical lattice was observed \cite{Chin2006}. By increasing the depth of the lattice potential near the Feshbach resonance, a superfluid-insulator transition was observed. They argued that the insulating state was the Mott insulator. 
In these experiments, it was argued that the usual single-band Hubbard model was no longer applicable, because the strength of the on-site interaction exceeded the gap between the lowest and the next-lowest bands. For detailed investigations of the experiments, accordingly, the effects of the higher bands have to be taken into account. 


Stimulated by these experiments, several theoretical studies on the superfluid-insulator transition were carried out \cite{Zhai2007,Chien0706,Moon0707}. However, both the correlation effects and the multiband effects have not yet been investigated well enough to discuss the transition from superfluid to Mott insulator. Precise studies including both effects are thus required. 


In this paper, we investigate the superfluid-insulator transition of interacting fermionic atoms in optical lattices, taking into account the multiband effects. For this purpose, we make use of a dynamical mean-field theory (DMFT) \cite{Georges1996}. This method enables us to treat local correlation effects precisely. We show that the multiband effects cause the transition from superfluid to Mott insulator. The phase diagram over the interaction from repulsive to attractive at half filling is presented. Features of each phase are discussed.


We consider that the fermionic atoms are in a periodic optical lattice potential. In the low-tunneling limit, each lattice potential is regarded as a harmonic one \cite{Zwerger2003,Hofstetter2006}. In this case, the lowest orbital is nondegenerate, while the next-lowest orbital is three-fold degenerate. 
We approximately model the system as follows. The degeneracy of the next-lowest orbital is neglected for simplicity. We introduce the effects of the lowest and the next-lowest orbitals. The hopping integrals between the same orbitals of the nearest neighbor sites are assumed to take the same value for both orbitals. 
According to the usual experiments, we assume that the system involves the same number of fermionic atoms in two different hyperfine states. These two hyperfine states are described as the pseudospin. 
The simplified model Hamiltonian we consider here reads 
\begin{eqnarray}
{\cal H} &=& \sum_{<i,j>\alpha\sigma}(t_{}^{}-\mu\delta_{i,j})
             c_{i\alpha\sigma}^{\dagger}c_{j\alpha\sigma}^{} 
            + \frac{D}{2}\sum_{i\sigma}(n_{i2\sigma}-n_{i1\sigma}) \notag \\
         &+& U^{}\sum_{i\alpha} n_{i\alpha\uparrow}n_{i\alpha\downarrow}
            + \sum_{i\sigma \sigma^{\prime}}
               (U^{\prime}-J \delta_{\sigma, \sigma'}) 
                n_{i1\sigma} n_{i2\sigma'},
\label{hami}
\end{eqnarray}
where $c_{i\alpha\sigma}$ is the fermionic annihilation operator for the state with pseudospin $\sigma$(=$\uparrow$,$\downarrow$) on orbital $\alpha$(=1, 2) in the $i$th lattice site and $n_{i\alpha\sigma}= c_{i\alpha\sigma}^{\dagger}c_{i\alpha\sigma}$. 
$t_{}^{}$ represents the hopping integral, $\mu$ the chemical potential, $D$ the splitting between the two orbitals. 
For the lowest and the next-lowest orbitals, the intraorbital on-site interaction $U$ between fermionic atoms, and the interorbital ones $U'$ and $J$ fulfill the relation $U'=J=U/2$. 
The Hamiltonian is known as a two-band Hubbard model. 
We assume that the intraorbital attractive interaction induces an $s$-wave superfluid state.


We investigate the ground state by using DMFT. In DMFT, the lattice model is mapped onto a single impurity model connected dynamically to a heat bath. We solve it self-consistently \cite{Georges1996}. This retains nontrivial local quantum fluctuations missing in conventional mean-field theories. There are various methods to solve the effective impurity model. In fact, the single-band attractive Hubbard model has been analyzed by some DMFT methods \cite{Keller2001,Capone2002,Garg2005,Toschi2005,Kyung2006}. We apply here the two-site DMFT method \cite{Potthoff2001}, which allows us to discuss the Mott transitions of orbitally degenerate lattice fermions qualitatively \cite{Ono2003,Koga2004}. Comprehensive investigation can be done, e.g., for the determination of phase diagrams. To study the superfluid of lattice fermions, we extend this method to the case when the superfluid order exists. The extension of the two-site DMFT to the multiband system is straightforward.


We use a semicircular density of states (DOS) $\rho_{}(\omega)=4/(\pi W_{})\sqrt{1-4(\omega/W_{})^{2}}$, where $W_{}$ is the band width. Since the hopping integrals are assumed to be independent of $\alpha$, $W_{}=4t$ and the DOS are the same for both bands.  
The chemical potential is set $\mu=U/2+U'-J/2$ so that particle-hole symmetry can be satisfied in both bands. In this case, the system is half filling. 
We neglect the density wave state which degenerates with the superfluid state in the symmetric case, because we focus on the superfluid state.
In the following, the hopping integral $t$ is used in units of energy.

\begin{figure}[htb]
\begin{center}
\includegraphics[scale=0.56]{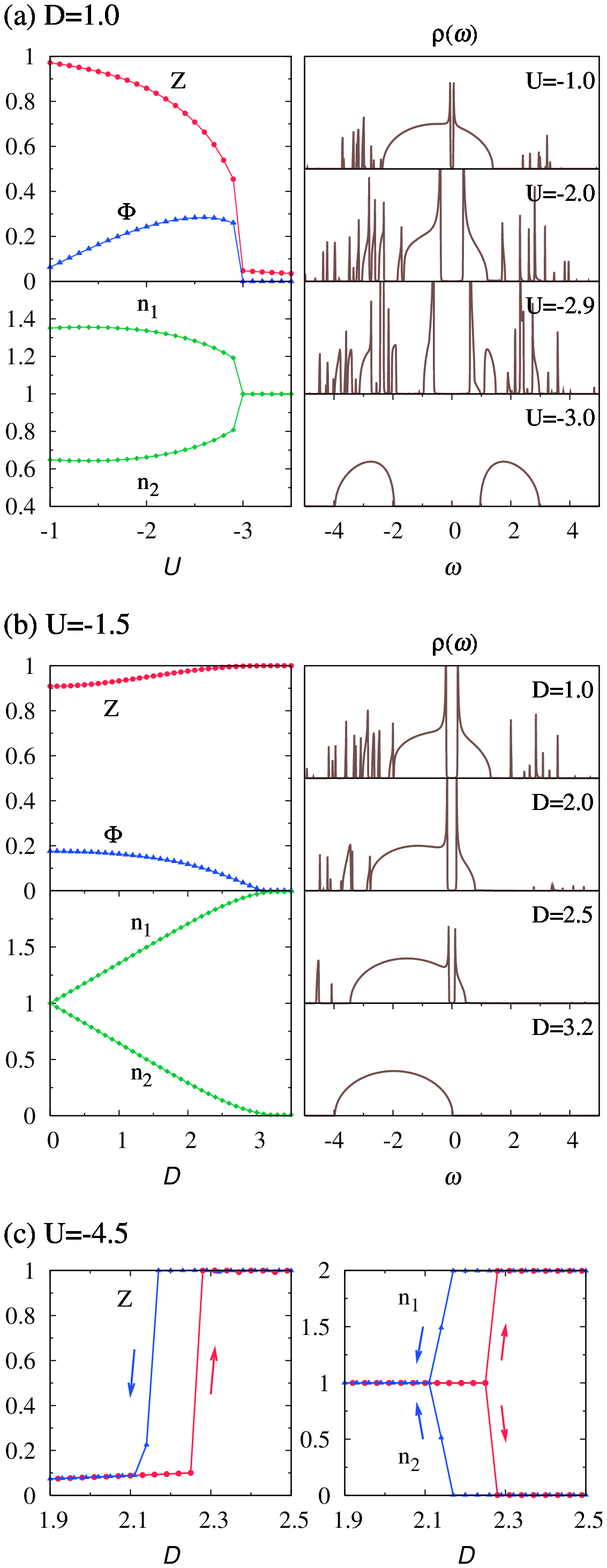}
\caption{The quasiparticle weight $Z$, the superfluid order parameter $\Phi$, and the filling of each band $n_{1},n_{2}$ for $U'=J=U/2$.
The DOS of the lowest band are shown in the right panels of (a) and (b). 
The results show three types of transitions: (a) the transition between the superfluid and the Mott insulator, (b) the transition between the superfluid and the band insulator, and (c) the transition between the band insulator and the Mott insulator. }
\label{DOS}
\end{center}
\end{figure}


We calculate the quasiparticle weight $Z$, the superfluid order parameter $\Phi=\langle c_{i\alpha\downarrow}c_{i\alpha\uparrow}\rangle$, and the filling of each band $n_{1}$, $n_{2}$ for $U'=J=U/2$ in the attractive $U$. $Z$ represents the coherent spectral weight of the Bogoliubov quasiparticle \cite{Garg2005}, which enables us to examine the Mott transition accurately. For half filling, $\Phi$ becomes independent of $i$ and $\alpha$ in the DMFT procedure. 
In Fig. \ref{DOS}(a), we show the results for $D=1.0$. 
As $|U|$ increases, $Z$ decreases to a small value close to zero at $U=-3.0$. $\Phi$ first increases and then decreases in $|U|>2.7$. At $U=-3.0$, $\Phi$ vanishes discontinuously. The significant decrease of $Z$ in the superfluid state is a manifestation of the strongly renormalized Bogoliubov quasiparticle. 
Because of the finite $D$, $n_1$ is larger than $n_2$ in $|U|<3.0$. As $|U|$ increases, $n_{1}$ decreases and $n_{2}$ increases, and both change to 1 suddenly at $U=-3.0$. 
The results indicate that the quantum phase transition from the superfluid to the Mott insulator takes place at $U=-3.0$ for $D=1.0$. The changes of the above quantities around $U=-3.0$ are rather discontinuous, suggesting that the transition may be of the first order. 
We also show the DOS of the lowest band for several values of $U$. We find a spectral gap around $\omega=0$, which is caused by the superfluid order. As $|U|$ increases, the spectral gap grows and the weight of the coherent part is reduced. The results are consistent with the decrease in the quasiparticle weight $Z$. 
Furthermore, the weight of the incoherent part increases steadily. 
At $U=-3.0$, the coherent part vanishes and the spectral gap by the superfluid order around $\omega = 0$ turns into the Hubbard gap. 
The results are consistent with each other concerning the quantum phase transition. Accordingly, we conclude that the transition from the superfluid to the Mott insulator takes place at $U=-3.0$. 
We emphasize that the Mott transition is caused by the multiband effects. In fact, the transition from the superfluid to the Mott insulator has not been obtained in the single-band attractive Hubbard model \cite{Garg2005,Toschi2005,Kyung2006}. 

In Fig. \ref{DOS}(b), we show $Z$, $\Phi$, $n_{1}$, and  $n_{2}$ as a function of $D$ for $U=-1.5$. The results are in noticeable contrast with those shown in Fig. \ref{DOS}(a). 
As $D$ increases, $Z$ increases to 1 and $\Phi$ decreases to 0 at $D=3.2$. We also observe that $n_{1}$ and $n_{2}$ change linearly from 1 in small $D$, and go to 2 and 0 at $D=3.2$, respectively. These changes are rather continuous, suggesting that the transition is of the second order. 
The results show that the quantum phase transition from the superfluid to the band insulator occurs at $D=3.2$ for $U=-1.5$. This conclusion is confirmed by the DOS of the lowest band. 
As $D$ increases, the spectral gap by the superfluid order becomes narrower and the weight of the incoherent part get smaller. For $D=3.2$, the spectral gap vanishes and the band insulator emerges. 

In Fig. \ref{DOS}(c), we show the results for $Z$ and $n_1$, $n_2$ in relatively strong attractive interaction $U=-4.5$. As $D$ increases, $Z$ changes from small value to 1, and the fillings change from $n_{1}= n_{2}=1$ to $n_{1}=2$ and $n_{2}=0$. The results imply that orbital fluctuations are suppressed and the direct transition from the Mott insulator to the band insulator takes place. 
In these quantities we observe the hysteresis, which is a proof of the first order transition. Judging from the present results and those in Fig. \ref{DOS}(a), we conclude that the Mott transition for the attractive interaction is of the first order. 
The transition point is determined at $D=2.3$ for $U=-4.5$, where the lowest energy is provided. 

\begin{figure}[htb]
\begin{center}
\includegraphics[scale=0.45]{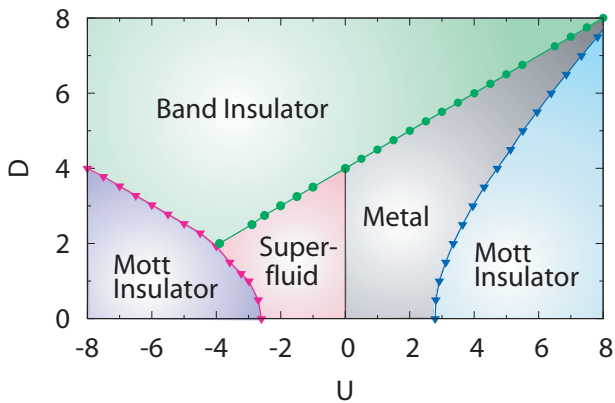}
\caption{Phase diagram for $U'=J=U/2$ at half filling. In the attractive side the bosonic fermion pair occupies either of the two orbitals in each site, while in the repulsive side the fermionic atoms occupy the respective two orbitals in each site.
} 
\label{PD}
\end{center}
\end{figure}


Changing $D$ and $U$ systematically, we investigate the quantum phase transition over the attractive and repulsive $U$. 
For the repulsive $U$, we obtain $(Z, n_1, n_2)=(1, 2, 0)$ and $(0, 1, 1)$. The former represents the band insulator and the latter the Mott insulator. We further obtain $0<Z<1$ and $0<n_2 \leq 1 \leq n_1<2$, which represent the metallic state. 
The results for $U'=J=U/2$ at half filling are summarized in the phase diagram shown in Fig. \ref{PD}. 
The band insulating phases smoothly connect between the repulsive and attractive sides. The transition line of the band insulator agrees well with the result by the conventional mean-field theory: $D=U/2+4$. Combining the results discussed in Fig. \ref{DOS}(b), we conclude that the band-insulator transition is of the second order. 
As the attractive interaction becomes strong, the superfluid state in the small $D$ region ($D<2$) changes directly into the Mott insulator owing to the strong renormalization of the correlation effects.  
In the appropriately large $D$ region ($2<D<4$), by contrast, the superfluid state first changes into the band insulator which is suitable for the fermion pair, and then the band insulator changes into the Mott insulator which is favorable in large $|U|$.


We now investigate the origin of the transition from the superfluid to the Mott insulator in another point of view. To this end, we map the model Hamiltonian at half filling to the effective boson model which is appropriate for the strong attractive $U$ \cite{com1}, 
\begin{eqnarray}
{\cal H_{\rm{Boson}}} = t_{\rm{eff}}^{B} \sum_{<i,j>} b_{i}^{\dagger}b_{j} +
                        U_{\rm{eff}}^{B} \sum_{i}n_{i}(n_{i}-1),
\label{hami_B}
\end{eqnarray}
where $b_{i}$ annihilates a bosonic fermion pair on the $i$th lattice site, $n_{i}=b_{i}^{\dagger}b_{i}$, $t_{\rm{eff}}^{B}$($=-2t^{2}/U$) represents the effective hopping integral of the fermion pair, and $U_{\rm{eff}}^{B}$($=-2U'+J-D$) denotes the effective interaction between two bosonic fermion pairs in the same lattice site. In deriving the effective boson model, we assume that $D$ is much smaller than $|U|$. 
Since the original model has only two orbitals, at most two bosonic fermion pairs occupy the same lattice site in the effective boson model. 
Under the condition for the lowest two orbitals: $U'=J=U/2(<0)$, we obtain $U_{\rm{eff}}^{B}>0$. The repulsive $U_{\rm{eff}}^{B}$ drives the superfluid state to the Mott insulator of bosonic fermion pairs. 
On the basis of the effective boson model, we discuss the Mott insulating states in the repulsive and attractive sides. 
Although the numerical results show no difference between them, the situations are different. 
For the Mott insulating state of the repulsive side, it is considered from the numerical results that the fermionic atoms in each site occupy respective two orbitals, which is consistent with the results obtained so far \cite{Koga2002,Werner2007}. 
For the Mott insulating state of the attractive side, by contrast, the bosonic fermion pairs occupy either of the two orbitals in each site. Therefore, the averaged fillings take the same value $n_1=n_2=1$, which is seemingly the same results as the repulsive side. 
As $D$ increases, orbital fluctuations in both sides play different roles. In the repulsive side, the metallic state sandwiched between the two insulating phases is stabilized by orbital fluctuations \cite{Koga2002,Werner2007}. 
In the strong attractive region, on the other hand, orbital fluctuations of fermion pairs are suppressed as mentioned before, leading to the direct transition between the two insulating states.

\begin{figure}[htb]
\begin{center}
\includegraphics[scale=0.7]{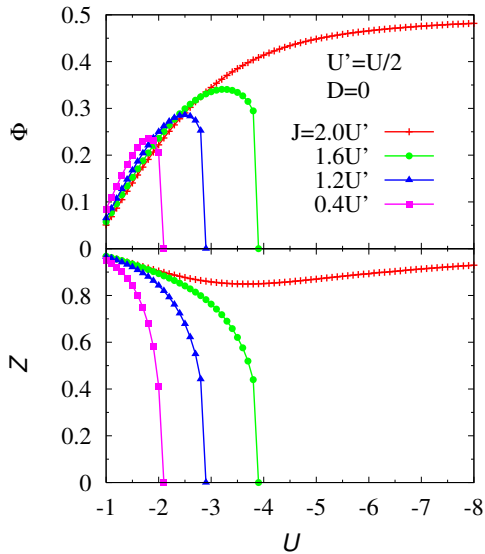}
\caption{The superfluid order parameter $\Phi$ and the quasiparticle weight $Z$ as a function of the attractive interaction $U$ for several values of $J$ in $U'=U/2$ and $D=0$. }
\label{Z}
\end{center}
\end{figure}

The expression \eqref{hami_B} suggests further that for $J=2U'$ and $D=0$ the effective interaction vanishes as $U_{\rm{eff}}^{B}=0$ and the Mott transition disappears. 
To confirm this suggestion, we calculate the quasiparticle weight $Z$ and the superfluid order parameter $\Phi$, releasing $J$ from the restriction $U'=J=U/2$. The results for several values of $J$ in $U'=U/2$ and $D=0$ are shown in Fig. \ref{Z}. We find that $Z \neq 0$ and $\Phi \neq 0$ for $J=2U'$. The results demonstrate that the Mott transition indeed disappears in agreement with the suggestion based on the effective boson model. On the other hand, the Mott transition occurs for $J \neq 2U'$. 
The agreement between the numerical results and the effective boson model indicates that the results based on the two-site DMFT and the description for the Mott insulator are reliable.


In summary, we have investigated the quantum phase transition of the fermionic atoms in optical lattices on the basis of a simplified model. Using two-site DMFT, we have shown that the transition from the superfluid to the Mott insulator at half filling is caused by the multiband effects. As a precursor of the Mott transition the renormalized superfluid state appears, which is characteristic of the interacting multiband fermionic atoms in optical lattices. 
The phase diagram among the superfluid, the Mott insulator, the band insulator, and the metallic state at half filling has been determined. 


We thank A. Yamamoto and M. Yamashita for valuable discussions. KI was supported by the Japan Society for the Promotion of Science. Numerical computations were carried out at the Supercomputer Center, the Institute for Solid State Physics, University of Tokyo.

\bibliography{apssamp}

\begin{thebibliography}{00}
\bibitem{com1} To obtain the effective model, we first make a particle-hole transformation in the higher orbital 2. We then map the obtained model to the effective boson model in the strong attractive $U$.
\end{thebibliography}


\end{document}